\documentclass[]{amsart}
\usepackage{amsmath,amssymb,amsfonts, bbm}
\usepackage{lmodern}
\usepackage[T1]{fontenc}
\usepackage[pdftex]{graphicx}

\newcommand{\CC}{\mathbb C}

\newcommand{\U}{\mathcal{U}}
\renewcommand{\u}{\mathfrak{u}}
\newcommand{\I}{\mathbbm{I}}
\newcommand{\1}{{\mathbf 1}}

\newcommand{\ad}{\operatorname{ad}}

\newcommand{\ntexp}{\operatorname{exp_{--}\!}}
\newcommand{\tr}{\operatorname{Tr}}

\newcommand{\diag}{\operatorname{diag}}

\newcommand{\T}{\operatorname{T}\!}

\renewcommand{\th}{\text{th}}

\newcommand{\HH}{\mathcal{H}}
\newcommand{\D}{\mathcal{D}}
\newcommand{\A}{\mathcal{A}}
\renewcommand{\L}{\mathcal{L}}
\renewcommand{\S}{\mathcal{S}}

\newcommand{\half}{\frac 12}

\newcommand{\dt}{\operatorname{d}\!t}

\newcommand{\eps}{\varepsilon}

\newcommand{\dd}[1]{\frac{\operatorname{d}}{\operatorname{d}\!#1}}
\newcommand{\length}[1]{\operatorname{Length}[#1]}

\newcommand{\dist}[2]{\operatorname{dist}(#1,#2)}

\begin{document}
\title[Geometry of quantum dynamics and optimal control for mixed states]{Geometry of quantum dynamics and optimal control for mixed states}
\date{\today}
\author{Ole Andersson and Hoshang Heydari}
\address{Department of Physics, Stockholm University, 10691 Stockholm, Sweden}
\email{olehandersson@gmail.com, hoshang@fysik.su.se}

\begin{abstract}
Geometric effects make evolution time vary for different evolution curves that connect the same two quantum states.
Thus, it is important to be able to control along which path a quantum state evolve
to achieve maximal speed in quantum calculations.
In this paper we establish fundamental relations between Hamiltonian dynamics and Riemannian structures on 
the phase spaces of unitarily evolving finite-level quantum systems. 
In particular, we show that the Riemannian distance between two density operators equals the infimum of the energy dispersions of all possible evolution curves connecting the two density operators. This means, essentially, that the evolution time is a controllable quantity.
The paper also contains two applied sections. First, we give a geometric derivation of the Mandelstam-Tamm estimate for the evolution time between two distinguishable mixed states. 
Secondly, we show how to equip the Hamiltonians acting on systems whose states are represented by invertible density operators with 
control parameters, and we formulate conditions for these that, when met, makes the Hamiltonians transport density operators along geodesics. 
\end{abstract}
\maketitle

\section{Introduction}
Geometric quantum mechanics is a branch of physics
that has received much attention lately. This is due in large part to the crucial role geometry plays in quantum information 
and quantum computing \cite{Pachos_etal1990,Zanardi_etal1999,Ekert_etal2000,Zanardi_etal2007,Rezakhani_etal2010,Sjoqvist_etal2012,Jones_etal2000,Falci_etal2000,Duan_etal2001,Recati_etal2002}.
The performance of a quantum computer relies greatly on the efficiency of its algorithms, and the ability to control which route an evolving state should take when joining two given ones. 
This paper concerns fundamental aspects of the latter issue.

A quantum system prepared in a pure state is usually modeled on a projective Hilbert space, and if the system is closed its state will evolve unitarily.
Aharonov and Anandan \cite{Anandan_etal1990} showed that for unitary evolutions there is a geometric quantity which, like Berry's phase \cite{Berry1984,Simon1983}, is independent of the particular Hamiltonian used to transport a pure state along a given route. 
More precisely, they showed that
the energy dispersion (i.e. $1/\hbar$ times the path integral of the energy uncertainty) of an evolving state equals the Fubini-Study length of
the curve traced out by the state. 
Using this, Aharonov and Anandan gave a new geometric interpretation of the time-energy uncertainty relation. 

Quantum computing and quantum information are theories developed for manipulating \emph{mixed} quantum states, i.e., statistical ensembles of pure states. Such states are usually represented by density operators.
Many metrics on spaces of density operators have been invented to capture various physical, mathematical, or information theoretical aspects of quantum mechanics.
In this paper we make use of a construction inspired by Montgomery \cite{Montgomery1991} to provide the spaces of isospectral density operators with Riemannian metrics, and we show that these metrics admit a generalization of the energy dispersion result of Aharonov and Anandan to evolutions of quantum systems in mixed states. 
Specifically, we show that the energy dispersion of an evolving mixed state is bounded from below by the length of the curve traced out by the density operator of the state. Furthermore, we show that 
every curve of isospectral density operators is generated by a Hamiltonian 
for which the energy dispersion equals the curve's length. The latter result allows us to express the distance between two mixed states in terms of a measurable quantity, and we use it to derive a time-energy uncertainty principle for mixed states. 

Generically, mixed states of finite-level quantum systems are represented by invertible density operators. 
In a concluding section we define control parameters for Hamiltonians of systems whose states are represented by invertible density operators, and give conditions for these that, when met, make the Hamiltonians transport density operators along geodesics. The section finishes with an application to almost pure qubit systems.

\section{Geometry of unitary quantum dynamics}
In this paper we consider finite dimensional quantum systems in mixed states that evolve unitarily.
They will be modeled on a Hilbert space $\HH$ of dimension $n$, and their states will be represented by density operators.
Evolving mixed states will be represented by \emph{curves} of density operators, all of which, for convenience, are assumed to be defined on an unspecified domain $0\leq t\leq \tau$. 
We write $\D(\HH)$ for the space of density operators on $\HH$.

\subsection{Riemannian structures on orbits of density operators}
A density operator that evolves unitarily remains in a single orbit of the left 
conjugation action of the unitary group $\U(\HH)$ of $\HH$ on $\D(\HH)$. 
The orbits of this action are in bijection with the set of possible spectra 
for density operators on $\HH$, where by the spectrum of a density 
operator with $k$-dimensional support we mean the \emph{nonincreasing} sequence 
$\sigma =\left(p_1,p_2,\dots,p_k\right)$ of its, not necessarily distinct, \emph{positive} eigenvalues. 
Throughout this paper we fix $\sigma$, and write $\D(\sigma)$ for the corresponding orbit.

To furnish $\D(\sigma)$ with a geometry, let $\L(\CC^k,\HH)$ be the space of linear maps from $\CC^k$ to $\HH$, 
equipped with the Hilbert-Schmidt inner product, and $P(\sigma)$ be 
the diagonal $k\times k$ matrix 
with diagonal $\sigma$.
Set $\S(\sigma)=\{\psi\in\L(\CC^k,\HH):\psi^\dagger \psi=P(\sigma)\}$,
and define $\pi:\S(\sigma)\to\D(\sigma)$ by $\pi(\psi)=\psi\psi^\dagger$.
The fibration $\pi$ is a principal bundle with left acting gauge group
$\U(\sigma)$ consisting of all unitaries in $\U(k)$ that commute with $P(\sigma)$.
(The action is $U\cdot\psi=\psi U^\dagger$.)
Moreover, 
the real part of the Hilbert-Schmidt product restricts to a gauge invariant Riemannian metric on $\S(\sigma)$:
\begin{equation*}
G(X,Y)=\half\tr(X^\dagger Y+Y^\dagger X).
\end{equation*}
We equip $\D(\sigma)$ with the unique metric $g$ that makes $\pi$ a Riemannian submersion.

The Lie algebra if the gauge group is
$\u(\sigma)$. It consists of all antiHermitian $k\times k$ matrices that commute with $P(\sigma)$. 
A connection for $\pi$ is given by the bundle of 
kernels of the mechanical connection form $\A$ on $\S(\sigma)$ defined by $\A_{\psi}=\I_{\psi}^{-1}J_{\psi}$,
where $\I_{\psi}:\u(\sigma)\to \u(\sigma)^*$ 
and $J_{\psi}:\T_{\psi}{\S(\sigma)}\to \u(\sigma)^*$ are the 
locked inertia tensor and moment map, respectively:
\begin{equation*}
\I_{\psi}\xi\cdot \eta=G(\psi\xi^\dagger,\psi\eta^\dagger),\qquad 
J_{\psi}(X)\cdot\xi=G(X,\psi\xi^\dagger).
\end{equation*}
Vectors tangent to $S(\sigma)$ are called horizontal if they are annihilated by $\A$, and a curve in $\S(\sigma)$ is called horizontal if all of its velocity vectors are
horizontal. Recall that for every curve
$\rho$ in $\D(\sigma)$ and every $\psi_0$ in the fiber over $\rho(0)$ there is a
unique horizontal curve in $\S(\sigma)$ that starts at $\psi_0$ and projects onto
$\rho$. This curve is the \emph{horizontal lift} of $\rho$ extending from $\psi_0$.

\subsection{A geometric uncertainty estimate}
Suppose $\hat A$ is an observable on $\HH$.
Let $X_A$ be the projection to $\D(\sigma)$ of  
the gauge invariant vector field $X_{\hat A}$ on $\S(\sigma)$ defined by
\begin{equation*}
X_{\hat A}(\psi)=\dd\eps\left[\exp\left(\frac{\eps}{i\hbar}\hat A\right)\psi\right]_{\eps=0}.
\end{equation*}
We say that $\hat{A}$ is parallel at a density operator $\rho$ if $X_{\hat{A}}$ is horizontal along the fiber over $\rho$,
and we say that $\hat{A}$ is parallel \emph{along a curve} $\rho$ if $\hat{A}(t)$ is parallel at $\rho(t)$ for every instant $t$.
The precision to which the value of $\hat A$ can be known is quantified by its uncertainty function
\begin{equation*}
\Delta A(\rho)=\sqrt{\tr(\hat A^2\rho)-\tr(\hat A\rho)^2}.
\end{equation*}
In \cite{GUR} the current authors have proven that $\Delta A$ is bounded from below by $\hbar$ times the
norm of the vector field $X_A$:  
\begin{align}
&\Delta A(\rho)\geq \hbar \sqrt{g(X_A(\rho),X_A(\rho))},\label{main ett}\\
&\Delta A(\rho)=\hbar\sqrt{g(X_A(\rho),X_A(\rho))}\text{ if $\hat{A}$ is parallel at $\rho$}.\label{main tva}
\end{align}
The main argument is the following.
For each $\psi$ in $\S(\sigma)$ there is a canonical identification between $\U(\sigma)$ and the fiber of $\pi$ containing $\psi$, namely $U\mapsto \psi U^\dagger$.
The metric on $\U(\sigma)$ obtained by pulling back $G$ via this identification is independent of $\psi$.
Restricted to $\u(\sigma)$ it is given by 
\begin{equation}\label{beta}
\xi\cdot\eta=\half\tr((\xi^\dagger\eta+\eta^\dagger\xi)P(\sigma)).
\end{equation}
Define the $\u(\sigma)$-valued field $\xi_A$ on $\D(\sigma)$ by $\pi^*\xi_A=\A\circ X_{\hat{A}}$,
and write $\xi_A^\bot$ for $\xi_A$ followed by projection onto the orthogonal complement of $i\1_k$ in $\u(\sigma)$.
Then 
\begin{equation*}
\Delta A^2=\hbar^2(g(X_A,X_A)+\xi_A^\bot\cdot\xi_A^\bot).
\end{equation*}
Now \eqref{main ett} follows from the observation that $\xi_A^\bot\cdot\xi_A^\bot\geq 0$, and 
\eqref{main tva} from the fact that $\xi_A(\rho)=0$ if $X_{\hat{A}}$ is horizontal along the fiber over $\rho$.
Note that for pure states, $\xi_A^\bot=0$ regardless of $\hat{A}$ since $\u(1)$ is spanned by $i\1_1$.

\subsection{Distance, geodesics, and energy dispersion}
The distance between two density operators with common spectrum $\sigma$ is defined as the infimum of the lengths of all curves in $\D(\sigma)$ that connects them. 
There is at least one such curve whose length equals the distance, since $\D(\sigma)$ is compact, 
and all such curves are geodesics. Moreover, horizontal lifting of curves is length preserving, $\pi$ being a Riemannian submersion, and a curve in $\D(\sigma)$ is a geodesic if and only if its horizontal lift is a geodesic in $\S(\sigma)$, see \cite{Hermann1960}.
Here we show that the distance between two density operators $\rho_0$ and $\rho_1$ with common spectrum $\sigma$ satisfies
\begin{equation}
\dist{\rho_0}{\rho_1}=\frac{1}{\hbar}\inf_{\hat H}\int_{0}^{\tau}\!\Delta H(\rho)\dt,\label{avstand}
\end{equation}
where the infimum is taken over all Hamiltonians $\hat H$ for which the following von Neumann equation is solvable: 
\begin{equation}
\dot\rho=\frac{1}{i\hbar}[\hat H,\rho]=X_H(\rho),\qquad \rho(0)=\rho_0,\quad \rho(\tau)=\rho_1.\label{von Neumann}
\end{equation}

The length of a curve $\rho$ in $\D(\sigma)$ is 
\begin{equation*}
\length{\rho}=
\int_{0}^{\tau}\!\sqrt{g(\dot\rho,\dot\rho)}\dt.\label{length}
\end{equation*}
If $ \dot\rho=X_H(\rho)$ for \emph{some} Hamiltonian $\hat H$, then, by \eqref{main ett}, the length of $\rho$ is a lower bound for the   
\emph{energy dispersion}: 
\begin{equation}
\length{\rho}\leq\frac{1}{\hbar}\int_{0}^{\tau}\!\Delta H(\rho)\dt.
\label{enekvation}
\end{equation}
There is a Hamiltonian $\hat H$ that 
generates a horizontal lift of $\rho$ because the unitary group of $\HH$ acts transitively on $\L(\CC^k,\HH)$.
For such a Hamiltonian we have equality in \eqref{enekvation} by \eqref{main tva}. Moreover, we can take $\rho$ to be a shortest geodesic. Then, 
\begin{equation*}
\dist{\rho_0}{\rho_1}=\frac{1}{\hbar}\int_{0}^{\tau}\!\Delta H(\rho)\dt.
\label{finalen}
\end{equation*}
Assertion \eqref{avstand} follows. We refer to \cite{GP} for a prescription how to produce a parallel transporting Hamiltonian from 
a given one, without affecting the evolution curve. 

\subsection{A time-energy uncertainty relation}
The Mandelstam-Tamm time-energy uncertainty relation \cite{Mandelstam_etal1945} provide a limit on the speed of dynamical evolution.
For systems prepared in pure states it implies that the minimum time it takes for a state to evolve to an orthogonal state is 
bounded from below by $\pi\hbar/2$ times the inverse of the average energy uncertainty of the system.
Recently, Jones and Kok \cite{Jones_etal2010, Zwierz2012} showed that the same inequality holds for mixed states, when 
orthogonality is replaced by \emph{distinguishability} \cite{Englert1996,Markham_etal2008}.
Their proof involves an estimate of the rate of change of the statistical distance between density operators.
Here we give a short geometric proof of this inequality.

Consider a quantum system with Hamiltonian $\hat{H}$, and suppose $\rho$ is a solution to \eqref{von Neumann}.
If $\rho_0$ and $\rho_1$ are distinguishable, then 
\begin{equation}
\langle \Delta H\rangle \tau\geq \frac{\pi\hbar}{2},\qquad \langle\Delta H\rangle=\frac{1}{\tau}\int_{0}^{\tau}\Delta H(\rho)\dt.
\label{energytime}
\end{equation}
To see this, let $\psi_0$ in $\pi^{-1}(\rho_0)$ and $\psi_1$ in $\pi^{-1}(\rho_1)$ be
such that $\dist{\rho_0}{\rho_1}=\dist{\psi_0}{\psi_1}$.
The operators $\rho_0$ and $\rho_1$ have orthogonal supports, being distinguishable, and the same is true for $\psi_0$ and $\psi_1$ 
since the the support of $\psi_0$ equals the support of $\rho_0$, and likewise for $\psi_1$ and $\rho_1$. A compact way to express this is
\begin{equation*}\label{vinkelrata}
\psi_0^\dagger\psi_1=0,\qquad \psi_1^\dagger\psi_0=0.
\end{equation*}
If we consider 
$\psi_0$ and $\psi_1$ elements in the unit sphere in $\L(\CC^k,\HH)$, they are a distance of $\pi/2$ apart. In fact, $\psi(t)=\cos(t)\psi_0+\sin(t)\psi_1$, with domain $0\leq t\leq \pi/2$, is a length minimizing unit speed curve from $\psi_0$ to $\psi_1$.
Consequently,  
\begin{equation}
\dist{\rho_0}{\rho_1}\geq \pi/2.\label{storre}
\end{equation} 
The relation \eqref{energytime} now follows from \eqref{enekvation} and  \eqref{storre}.
Also note that the estimate \eqref{storre} cannot be improved. Direct computations yield $\psi^\dagger\psi=P(\sigma)$ and $\psi^\dagger\dot\psi=0$. Therefore, $\psi$ is a \emph{horizontal} curve \emph{in} $\S(\sigma)$, and hence \eqref{storre} is, in fact, an equality.  From this it also follows that the estimate \eqref{energytime} is saturated by any parallel Hamiltonian that generate $\rho=\psi\psi^\dagger$.

\section{Optimal Hamiltonians for mixed states of full rank}
Generically, the number of independent kets in a mixed state equals the dimension of the Hilbert space.
Such mixed states are represented by invertible density operators. 
From now on, we assume that the density operators in $\D(\sigma)$ are invertible.

\subsection{Lie algebra controlled Hamiltonians}
To achieve optimal computational speed in quantum computers it is desirable that the Hamiltonians transport states along shortest possible paths. Here we classify the Hamiltonians that transport an invertible density operator $\rho_0$ along geodesics, and we provide conditions for control parameters of these Hamiltonians
that, when satisfied, makes the Hamiltonians transport a $\psi_0$ in the fiber over $\rho_0$ along horizontal geodesics.

As control space we choose the matrix Lie algebra $\u(n)$, equipped with the metric given by \eqref{beta}.
For each curve $\xi$ in $\u(n)$ we define $\hat{H}_\xi$ by
\begin{equation}\label{Ham}
\hat{H}_\xi=i\hbar \psi_0 P(\sigma)^{-1/2}\ntexp\left(\int_0^t\!\xi\dt\right)\xi \ntexp\left(\int_0^t\!\xi\dt\right)^\dagger P(\sigma)^{-1/2}\psi_0^\dagger.
\end{equation}
where $P(\sigma)^{-1/2}$ is the diagonal $n\times n$ matrix whose $j^{\th}$ diagonal entry is $1/\sqrt{p_j}$, and $\ntexp$ is the negative time-ordered exponential. Also, let $\psi_\xi$ be the solution to the Schr\"{o}dinger equation of $\hat{H}_\xi$ extending from $\psi_0$.
In the next section we show that every curve $\rho$ extending from $\rho_0$ equals $\psi_\xi\psi_\xi^\dagger$, for some curve $\xi$ in $\u(n)$,
and that $\psi_\xi$ is a geodesic if and only if $\dot{\xi}=\ad^*_\xi\xi$, where $\ad^*_\xi\xi\cdot\eta=\xi\cdot [\xi,\eta]$. Furthermore, we show that $\psi_\xi$ is horizontal if and only if $\xi$ is contained in the orthogonal complement $\u(\sigma)^\bot$ of $\u(\sigma)$ in $\u(n)$.

\subsection{Evolution operators that generate horizontal geodesics}\label{Hamiltonians whose Schrödinger equations have horizontal geodesic solutions}
The group $\U(\HH)$ acts freely and transitively on $\S(\sigma)$ from the left, 
and the metric on $\U(\HH)$ obtained by declaring the diffeomorphism $U\mapsto U\psi_0$ an isometry is the left invariant metric $X\cdot Y=\tr((X^\dagger Y+Y^\dagger X)P(\sigma))$.
An evolution curve $U\psi_0$ is a geodesic in $\S(\sigma)$ if and only if $U$ is a geodesic in $\U(\HH)$ extending from the identity operator.
The second order geodesic equation in $\U(\HH)$ can be reduced to a first order equation in $\u(n)$ as follows. 
Define an isomorphism $\phi:\U(n)\to \U(\HH)$ by 
$\phi(U)=\psi_0 P(\sigma)^{-1/2} UP(\sigma)^{-1/2}\psi_0^\dagger$. Equip $\U(n)$ with the metric that makes $\phi$ an isometry, and for each $\xi$ in $\u(n)$ write $\chi_\xi$ for the 
vector field on $\U(\HH)$ made up of left translates of $d\phi(\xi)$.
For a given curve $U$ in $\U(\HH)$ define a curve $\xi$ in $\u(n)$ by $\dot U=\chi_\xi(U)$.
The Hamiltonian associated with $\xi$, that transport $\psi_0$ along $U\psi_0$ in $\S(\sigma)$, is $\hat{H}_\xi$ given by \eqref{Ham}.
Now, $U$ is a geodesic 
if and only if $\xi$ satisfies the Arnold-Euler equation $\dot{\xi}=\ad^*_\xi\xi$, see \cite{Arnold_1966} and Appendix \ref{CAEE}.
Moreover, the inclusion of $\U(\sigma)$ in $\U(n)$ is an isometric embedding, when the former is equipped with the 
left invariant metric determined by \eqref{beta}. A straightforward verification shows that $U\psi_0$ is horizontal if and only if $\xi$ is contained in $\u(\sigma)^\bot$, see Appendix \ref{CAEE}.

\subsection{Geodesic orbit spaces and almost pure states}\label{two eigenvalues}
If $\sigma$ contains precisely two different, possibly degenerate, eigenvalues, every geodesic 
in $\D(\sigma)$ is generated by a time independent Hamiltonian. 
This since $\ad_\xi^*\xi=0$ holds for every $\xi$ in $\u(\sigma)^\bot$.
To see this let $\eta$ be any element in $\u(n)$, and write
\begin{equation*}
\xi=\begin{bmatrix} 0 & \xi_{12}\\ -\xi_{12}^\dagger & 0\end{bmatrix},
\qquad
\eta=\begin{bmatrix} \eta_{11} & \eta_{12}\\ -\eta_{12}^\dagger & \eta_{22}\end{bmatrix}.
\end{equation*}
Then
\begin{equation*}
\xi\cdot [\xi,\eta]
=\half\left(p_1\tr[\xi_{12}\xi_{12}^\dagger,\eta_{11}] + p_2\tr[\xi_{12}^\dagger\xi_{12},\eta_{22}]\right)=0
\end{equation*}
since commutators have vanishing trace. (The corresponding result does not hold if $\sigma$ contains at least three distinct eigenvalues.) Another way to put this is to say that $\D(\sigma)$ is a \emph{geodesic orbit space}, i.e., a Riemannian homogeneous space 
in which each geodesic is an orbit of a one-parameter subgroup of its isometry group.

By an \emph{almost pure state} we mean a mixture of two pure quantum states
in which one state is present in greater proportion than the other.
Here we apply the above results to almost pure qubit systems.

Two independent qubits are modeled by the standard basis elements $e_1$ and $e_2$ in $\CC^2$.
Consider an ensemble of qubits prepared so that the proportion of qubits in state $e_j$ is $p_j$, where $p_1>p_2$.
The initial state of the ensemble is represented
by the density operator $\rho_0=\diag(p_1,p_2)$. 
Chose $\psi_0=\diag(\sqrt{p_1},\sqrt{p_2})$ in the fiber over $\rho_0$, 
and let $\xi$ be an arbitrary constant curve in $\u(\sigma)^\bot$:
\begin{equation*}
\xi(t)=\begin{bmatrix} 0 & \eps e^{i\theta}\\ -\eps e^{-i\theta} & 0\end{bmatrix},\qquad \eps>0,\quad 0\leq t\leq 1.
\end{equation*}
The solution $\psi_\xi$ to the Schr\"odinger equation of $\hat{H}_\xi=i\hbar\xi$ that extends from $\psi_0$ is a horizontal geodesic,
and the projection $\rho=\psi_\xi\psi_\xi^\dagger$ is a geodesic extending from $\rho_0$. Explicitly, 
\begin{equation*}
\rho(t)=\begin{bmatrix} p_1\cos^2\eps t+p_2\sin^2\eps t & e^{i\theta}(p_2-p_1)\cos\eps t\sin\eps t\\ e^{-i\theta}(p_2-p_1)\cos\eps t\sin\eps t & p_1\sin^2\eps t+p_2\cos^2\eps t\end{bmatrix}.
\end{equation*}
The curve $\rho$ is a shortest geodesic between its end points provided that $\eps$ is small enough, and $\dist{\rho(0)}{\rho(1)}=\eps$.

\section{Conclusion}
The classic time-energy uncertainty relation by Mandelstam and Tamm implies that the evolution time between 
two distinguishable mixed states is bounded from below by a factor which is inversely proportional to the average energy uncertainty.
In this paper we have shown that there is a fundamental relation between the length of evolution curves of mixed states, as measured by a specific Riemannian metric,
and the energy dispersions of the Hamiltonians that generate the evolution curves. Our work thus indicates that the evolution time estimate derived from the Mandelstam-Tamm relation has a purely geometric origin. In fact, we have provided a geometric derivation of the same estimate.

In quantum computing it is desirable to have greatest possible control over the evolution of states. This to achieve maximum computational speed.
Generically, mixed states of finite-level quantum systems are represented by invertible density operators.
In the paper's concluding section, we have focused on quantum systems whose states are represented by invertible density operators. There we have described how Hamiltonians acting on such systems can be equipped with control parameters, and we have provided conditions for these that, if met, guarantees that the Hamiltonians transport density operators along geodesics.

\section*{acknowledgement}
HH acknowledges financial support by the Swedish Research Council (VR), grant number 2008-5227.

\appendix

\section{Constrained Arnold-Euler equation}\label{CAEE}
Suppose the density operators with spectrum $\sigma$ are invertible. Fix $\psi_0$ in $\S(\sigma)$, and identify $\U(n)$ and $\U(\HH)$ according to the isomorphism $\phi:\U(n)\to\U(\HH)$ given by \begin{equation*}
\phi(U)=\psi_0 P(\sigma)^{-1/2}U P(\sigma)^{-1/2}\psi_0^\dagger.
\end{equation*} 
Then $\U(n)$ acts freely and transitively from the left on $\S(\sigma)$. Thus $U\mapsto\phi(U)\psi_0$ is a diffeomorphism from $\U(n)$ to $\S(\sigma)$.
Let $Y_\xi$ be the push-forward of the one-parameter family of left invariant vector fields on $\U(n)$ generated by a curve $\xi$ in $\u(n)$:
\begin{equation*}
Y_\xi(\phi(U)\psi_0)=\psi_0 P(\sigma)^{-1/2}U\xi P(\sigma)^{1/2}.
\end{equation*}
We assert that the integral curves of $Y_\xi$ are horizontal geodesics if and only if $\xi$ is contained in $\u(\sigma)^\bot$ and $\dot{\xi}=\ad_\xi^*\xi$. The latter equation is called the Arnold-Euler equation \cite{Arnold_1966}.

The mechanical connection is such that a tangent vector is horizontal if and only if it is orthogonal to the fibers of $\pi$.
The tangent space at $\phi(U)\psi_0$ of the fiber of $\pi$ is spanned by the vectors $\phi(U)\psi_0\eta^\dagger$, where $\eta$ run through the matrices in $\u(\sigma)$. A straightforward computation yields
\begin{equation*}
G(Y_\xi(\phi(U)\psi_0),\phi(U)\psi_0\eta^\dagger)=-\half\tr((\xi^\dagger\eta+\eta^\dagger\xi)P(\sigma)).
\end{equation*}
We conclude that $Y_\xi$ is horizontal if and only if $\xi$ is contained in $\u(\sigma)^\bot$.
 
Let $\psi$ be an integral curve of $Y_\xi$.
Its covariant derivative satisfies 
\begin{equation*}
\nabla_\psi\dot \psi=X_{\dot\xi}(\psi)+\nabla_{X_\xi}X_{\xi}(\psi).
\end{equation*}
Moreover, by the Kozul formula \cite[Prop 2.3]{Kobayashi_etal1996},
\begin{equation*}
\begin{split}
2G(\nabla_{X_\xi}X_\xi,X_\eta)
&=X_{\xi}G(X_{\xi},X_{\eta})+X_{\xi}G(X_{\eta},X_{\xi})\\
&\qquad\qquad-X_{\eta}G(X_{\xi},X_{\xi})-G(X_\xi,[X_\xi,X_\eta])\\
&\qquad\qquad+G(X_\xi,[X_\eta,X_\xi])+G(X_\eta,[X_\xi,X_\xi])\\
&=-\xi\cdot[\xi,\eta]+\xi\cdot[\eta,\xi]\\
&=-2\ad_\xi^*\xi\cdot\eta\\
&=-\,2G(X_{\ad_\xi^*\xi},X_\eta)
\end{split}
\end{equation*}
for every $\eta$ in $\u(n)$. Accordingly, $\nabla_\psi\dot \psi= X_{\dot\xi-\ad_\xi^*\xi}(\psi)$. Thus, $\psi$ is a geodesic if and only if $\dot\xi=\ad_\xi^*\xi$.

\end{document}